\documentclass[aps,amsmath,showpacs,showkeys]{revtex4}
\usepackage{epsf}
\newcommand{\fr}[2]{\frac{#1}{#2}}
\newcommand{\Ref}[1]{(\ref{#1})}
\newcommand{\be}{\begin{equation}}
\newcommand{\ee}{\end{equation}}
\newcommand{\bn}{\begin{eqnarray}}
\newcommand{\en}{\end{eqnarray}}
\newcommand{\bd}{\begin{displaymath}}
\newcommand{\ed}{\end{displaymath}}
\newcommand{\bnn}{\begin{eqnarray*}}
\newcommand{\enn}{\end{eqnarray*}}
\newcommand{\adb}{\allowdisplaybreaks }
\newcommand{\bs}{\begin{subequations}}
\newcommand{\es}{\end{subequations}}
\newcommand{\non}{\nonumber}
\newcommand{\ls}{\lambda_s}
\newcommand{\lso}{\lambda_{s1}}
\newcommand{\lst}{\lambda_{s2}}

\begin{document}

\title{Zeta-function approach to Casimir energy with singular potentials}

\author{Nail R. Khusnutdinov\footnote{e-mail: nail@kazan-spu.ru}%
\footnote{On leave from Department of Physics, Kazan State
Pedagogical University, Mezhlauk 1, Kazan 420021, Russia}}

\address{Departamento de F\'{\i}sica, Universidade Federal da
Para\'{\i}ba, Caixa Postal 5008, CEP 58051-970 Jo\~ao Pessoa, Pb,
Brazil}

\date{\today}

\begin{abstract}
In the framework of zeta-function approach the Casimir energy for
three simple model system: single delta potential, step function
potential and three delta potentials is analyzed. It is shown that
the energy contains contributions which are peculiar to the
potentials. It is suggested to renormalize the energy using the
condition that the energy of infinitely separated potentials is
zero which corresponds to subtraction all terms of asymptotic
expansion of zeta-function. The energy obtained in this way obeys
all physically reasonable conditions. It is finite in the
Dirichlet limit and it may be attractive or repulsive depending on
the strength of potential. The effective action is calculated and
it is shown that the surface contribution appears. The
renormalization of the effective action is discussed.
\end{abstract}
\pacs{11.10.Gh; 11.10.-z; 03.70.+k} \keywords{zeta-function;
Casimir effect; Zero-point energy; Renormalization; Heat-kernel
coefficients}

\maketitle

\section{Introduction}

Recently there was great interest to Casimir effect for
semi-transparent boundaries
\cite{BorKirVas99,Sca99,GraJafKheQuaScaWei02,Mil03,BorVas04,Mil04}.
It was suggested to consider the potential with zero support
instead of the rigid Dirichlet condition. This potential models
the semi-transparent boundary condition. The space is not divided
into separate parts by Dirichlet boundaries and some modes of
field may go through the boundary. Firstly, this kind of
calculations were made in Refs.
\cite{MamTru82,BorHenRob92,MosTru97}. It was claimed that in the
limit of infinite strength of potential, which was called
Dirichlet limit \cite{GraJafKheQuaScaWei02}, the energy is
divergent and it does not coincide with that obtained for
Dirichlet boundary condition. The authors of
\cite{GraJafKheQuaScaWei02} noted that in framework of QED it is
impossible to obtain finite result. This divergence appears in the
energy only. The force, which is the derivative of energy with
respect the position of plate, is finite \cite{Mil04} in the
Dirichlet limit.

The same problem was emphasized in Ref. \cite{BorVas04}. Singular
potential brings additional surface contributions to heat kernels
coefficients \cite{BorVas99} and to the effective action and there
is no universal method to make unambiguous renormalization
procedure and fix all parameters. Necessity to consideration of
the surface contributions to energy in framework of bag model was
noted in Ref. \cite{Mil80,BlaVisWipf88}. From the point of view of
field theory the delta-like potential may be considered as brane
theory and the parameter of potential strength is regarded as
brane's mass \cite{GolWis01}. In framework of quantum field theory
this parameter has to be renormalized, too. Renormalization group
equation for brane with co-dimension greater then unit was
obtained in Ref. \cite{GolWis01}. It was noted that for
renormalization we have to introduce into theory not merely brane
mass but also brane tension \cite{GeoGraHai01,BorVas04} and others
parameters \cite{GolWis01} the number of which depends on the bulk
action. There is another analogy of system under consideration
with field theory in curved space-time with singular scalar
curvature. For example, in space-time of short-throat wormhole
\cite{KhuSus02} the space-time is everywhere flat except the
throat where the scalar curvature is singular. The strength of the
delta-potential corresponds to non-conformal coupling constant
which has to be renormalized too \cite{Odi91}.

In this paper we reexamine this problem in framework of
zeta-regularization approach considering in details some simple
model system in Sec. \ref{Sec:Zeta} namely, single delta
potential, step function potential and three delta potentials. The
main features of these models are summarized in Sec.
\ref{SecRenormalization}. It was noted that the singularities
appeared in the limit of infinite strength are connected with
brain (potential) itself. There we suggest a method to extract
physically reasonable expression for energy which obey all
conditions we need. It is shown that the same result may be
obtained by Lukosz renormalization procedure. The Casimir energy
may possesses the maximum and the Casimir force may be attractive
as well as repulsive. Close analogy with "surface energy"\/ is
noted. The effective action in framework of zeta-regularization
for $\phi^4$ theory with delta potential is considered in Sec.
\ref{Sec:Eff}. Recently similar approach was used by Toms in Ref.
\cite{Tom05} for free scalar field. The same ambiguous of
renormalization procedure in surface term \cite{BorVas04} is
recovered. We observe that this ambiguous is connected with some
surface quantity peculiar to the potential (brane) itself. By
using approach of Sec. \ref{SecRenormalization} we obtain
physically reasonable result which is coincides with that obtained
in Sec. \ref{SecRenormalization}. We note that the Lukosz
renormalization procedure takes off all singularities of effective
action.

\section{The zeta-regularization approach}\label{Sec:Zeta}

Let us consider the massive scalar field in $N+1$ dimensional
Minkowski space-time with potential $V$ which depends on single
coordinate $x$, only. In the framework of zeta-regularization
approach \cite{DowCri76,BlaVisWipf88,BorEliKirLes97} the
regularized energy is defined by following expression
\be\label{Edef}
E^{(N)}(s) = \fr 12 \mu^{2s} \zeta(s-\fr 12,D_N)
\ee
where
\bd
\zeta(s,D_N) = \sum_{(n)} \omega_{(n)}^{-2s}
\ed
is the zeta-function of operator
\bd
D_N = -\triangle_N + m^2 + V
\ed
with eigen-values $\omega_{(n)}^2$. Parameter $\mu$ with dimension
of mass has introduced to keep right dimension of the energy. In
accordance with Ref. \cite{BorEliKirLes97} one defines the Casimir
energy by relation
\be\label{Emain}
E_c^{(N)} = \lim_{s\to 0} (E^{(N)}(s) - E^{(N)}_{div}(s)),
\ee
where
\bd
E^{(N)}_{div}(s) = \lim_{m\to \infty} E^{(N)}(s) =
\left(\fr{\mu}{m}\right)^{2s} \fr 1{2(4\pi)^{N/2}}
\sum_{n=0}^{N+1} B_{\fr n2} m^{N+1-n} \fr{\Gamma (s+
\fr{n-N-1}{2})}{\Gamma (s-\fr 12)}
\ed
is the divergent part of energy, where $B_\alpha$  are the heat
kernel coefficients of operator $D_N$. This expression obeys to
the physical reasonable condition
\be\label{condition}
\lim_{m\to \infty}E_c^{(N)} =0.
\ee

The spectrum of this operator is numerated by one discrete number
$n$, and continuous numbers $\mathbf{k}_{N-1} \in (-\infty,
+\infty)$
\bd
\omega^2 = k_n^2 + \mathbf{k}_{N-1}^2 + m^2.
\ed
By integrating the definition of zeta function over continuous
variables $\mathbf{k}_{N-1}$ we obtain the relation
\be\label{zetaN}
\zeta(s - \fr 12,D_N) = \fr 1{(4\pi)^{\fr{N-1}2}} \fr{\Gamma
(s-\fr N2)}{\Gamma (s-\fr 12)} \zeta(s - \fr N2,D_1)
\ee
by using which we have to solve now the one dimensional problem
only. We note that the using above relation in Eq. \Ref{Edef}
gives the energy per unit square of plate, that is surface density
of energy. To calculate the zeta-function we will use approach
\cite{BorEliKirLes97} in framework of which the zeta-function is
represented in the form below
\be\label{zeta1}
\zeta(s,D_1) = \fr{\sin (\pi s)}{\pi} \int_m^\infty dk (k^2 -
m^2)^{- s} \fr\partial{\partial k} \ln \Psi(ik,R,R')
\ee
where the spectrum of energy $k^2 = m^2 - \omega^2$ is found from
relation $\Psi (k,R,R')=0$. Let us proceed to consideration of
different forms of potential.

\subsection{Single singular potential $V=\ls \delta (x)$}\label{Sec:Single}

We consider a singular potential $V = \ls \delta (x)$ and two
$N-1$ dimensional Dirichlet plates at points $x=R$ and $x=-R'$.
Because of Eq. \Ref{zeta1} we need to consider the imaginary
energies only. Let us denote $m^2 - \omega^2 = k^2$ and consider
the one dimensional problem in imaginary axis $k\to ik$. In this
case the eigen-problem equation reads
\bd
\left[-\fr{\partial^2 }{\partial x^2}  + k^2  +
\ls\delta(x)\right]\phi = 0.
\ed
In two domains $x\in (0,R)$ and $x\in (-R',0)$ we obtain the
general solution of this equation
\bnn
\phi^+ &=& C_1^+ e^{kx} + C_2^+e^{-kx},\\
\phi^- &=& C_1^- e^{kx} + C_2^-e^{-kx}.
\enn
Four constants $C_1^\pm$ and $C_2^\pm$ obey to four homogenous
equations
\bnn
\phi^+(0) &=& \phi^-(0),\\
\phi^+{}'(0) &=& \phi^-{}'(0) + \ls \phi^+(0),\\
\phi^+(R) &=& 0,\\
\phi^-(-R') &=& 0.
\enn
The solution of this system exists if and only if the determinant
of the system equals to zero. This determinant is the function
$\Psi(ik,R,R')$ which is used in Eq. \Ref{zeta1}:
\be\label{PsikRr}
\Psi(ik,R,R') = \fr 1k \left[e^{-kR'} A' + e^{kR'} A\right],
\ee
where
\bnn
A' &=& \fr 1{k} \left[\ls e^{kR}+ \left(2k - \ls \right)
e^{-kR}\right],\\
A &=& \fr 1{k} \left[\ls e^{-kR} - \left(2k + \ls \right)
e^{kR}\right].
\enn
It is easy to see from Eq. \Ref{PsikRr} by equating
$\Psi(ik,R,R')$ to zero that the boundary states appear if
\be\label{BoundIneq}
\ls < - \fr{R+R'}{RR'}.
\ee
This boundary state is localized in the potential (brane) and in
the case of whole space $R=R'\to \infty$ it has the form below
\cite{MamTru82}:
\bd
\phi (x) = C e^{-|\ls x|}.
\ed
Because the possible boundary states have already incorporated in
Eq. \Ref{zeta1} we assume arbitrary sign of $\ls$. Nevertheless we
suppose that the inequality
\be
\ls > -2 m
\ee
is valid. In opposite case we can not use the theory in present
form (see Ref. \cite{MamTru82}).

To find the heat kernel coefficients which are defined as
asymptotic expansion over mass,
\bd
\zeta(s,D_1)^{as} = \fr{1}{\sqrt{4\pi}} \fr 1{\Gamma(s)}
\sum_{n=0}^\infty B_{\fr n2} m^{-2 s + 1 -n}\Gamma(s - \fr 12 +
\fr n2),
\ed
we take the asymptotic expansion of zeta function \Ref{main} over
$k\to \infty$. One has
\be
\fr\partial{\partial k} \ln \Psi(k,R,R')^{as} = R+R' - \fr 1k -
\fr{\ls}{k(2k+\ls)}= R+R' - \fr 1k + \sum_{l=1}^\infty (-1)^l
\fr{\ls^l}{2^lk^{l+1}}.\label{AsExtr}
\ee
By using this expression we obtain heat kernel coefficients
\bn\label{hkcdelta}
B_0 &=& R+R',\ B_{\fr 12} = -\sqrt{\pi},\non\\
B_{n} &=& - \left(\fr\ls 2\right)^{2n-1} \fr{\sqrt{\pi}}{\Gamma
(n+\fr 12)},\\
B_{n+\fr 12} &=& \left(\fr\ls 2\right)^{2n} \fr{\sqrt{\pi}}{n!},\
n=1,2,3,\ldots .\non
\en

Then, using Eqs. \Ref{zetaN} and \Ref{zeta1} we arrive at the
following expression for Casimir energy
\be\label{EN}
E_c^{(N)}[\ls,R,R'] = -\fr 1{2(4\pi)^{\fr{N}2}} \fr{1}{\Gamma (\fr
N2 + 1)} \int_m^\infty dk (k^2 - m^2)^{N/2} {\cal
Z}_N[k,\ls,R,R'],
\ee
where
\be\label{Ren}
{\cal Z}_N[k,\ls,R,R'] = \fr\partial{\partial k} \ln \Psi(k,R,R')
- (R+R') + \fr 1k - \sum_{l=1}^N (-1)^l \fr{\ls^l}{2^lk^{l+1}}.
\ee
Let us consider some limiting cases. In massless limit $m\to 0$
the energy is divergent
\be\label{Eas}
\left.E_c^{(N)}[\ls,R,R']\right|_{m\to 0} \approx \fr{
(-1)^N}{(4\pi)^{\fr{N+1}2}} B_{\fr{N+1}{2}} \ln \fr{m}{\ls}
\ee
in agreement with Ref. \cite{BorKirVas99}. The energy in whole
space ($R=R'\to \infty$),
\bn\label{Ediv}
&&E_c^{(N)}[\ls,R\to\infty,R'\to\infty] = \fr 1{2(4\pi)^{\fr{N}2}}
\fr{1}{\Gamma (\fr N2 + 1)} \int_m^\infty dk (k^2 -
m^2)^{N/2}\left[\fr{\ls}{k(2 k + \ls)} +
\sum_{l=1}^N (-1)^l \fr{\ls^l}{2^lk^{l+1}}\right]\\
&=&\fr{(-1)^{N+1}m^N}{4(4\pi)^{N/2}}\left(\fr{\ls}{2m}\right)^{N+1}
\left\{\fr{\sqrt{\pi}}{\Gamma[\fr{N+3}{2}]} {}_2F_1\left[1,\fr
12;\fr{N+3}{2};\left(\fr{\ls}{2m}\right)^2\right] - \fr{\ls}{2m}
\fr{1}{\Gamma[\fr{N+4}{2}]}{}_2F_1\left[1,1;\fr{N+4}{2};
\left(\fr{\ls}{2m}\right)^2\right]\right\},\non
\en
is finite but it is ill defined in the Dirichlet limit,
$\ls\to\infty$. Here the ${}_2F_1$ is the hypergeometric function.
The leading divergent term in this limit coincides with that in
Eq. \Ref{Eas}. The energy is well defined in the limit $\ls\to 0$.
In this case
\bd
{\cal Z}_N[k,\ls\to 0,R,R']= \fr{2(R+R')}{e^{2k(R+R')}-1}
\ed
which corresponds to the Casimir energy for two Dirichlet plates
at points $-R'$ and $R$.

Let us consider the energy for particular dimensions $N=1,3$ in
manifest form. Then for $N=1$ we obtain
\bd
E_c^{(1)}[\ls,R,R'] = - \fr{m}{2\pi} \int_1^\infty dx (x^2 -
1)^{1/2} {\cal Z}_1[mx,\ls,R,R'],
\ed
where
\bnn
&&{\cal Z}_1[k,\ls,R,R']\\
&=& \left\{8 k^3 (R+R') + 2k \ls \left[-2 + e^{2kR} +e^{2kR'} +
2kR'(e^{2kR}-1) + 2kR (e^{2kR'}-1)\right] +
(e^{2kR}-1)(e^{2kR'}-1)
\ls^2\right\}\non\\
&\times&\left\{2k^2\left[2k(e^{2k(R+R')}-1) +
(e^{2kR}-1)(e^{2kR}-1)\ls \right]\right\}^{-1}. \non
\enn
In the limit $R=R'\to\infty$
\bd
E_c^{(1)}[\ls,R\to\infty,R'\to\infty] = -\fr{m}{4\pi} \left(\pi -
\fr{\ls}{m} - \sqrt{4-\fr{\ls^2}{m^2}} \arccos
\fr{\ls}{2m}\right).
\ed
This expression  coincides exactly with that obtained in
Ref.\cite{GraJafKheQuaScaWei02}. It is ill defined in the
Dirichlet limit $\ls\to\infty$:
\bd
E_c^{(1)}[\ls\to\infty,R\to\infty,R'\to\infty] \approx
\fr{\ls}{4\pi} (1-\ln \fr{\ls}{2m}) - \fr m4.
\ed

The force acting on the plate coincides with that obtained in Ref.
\cite{Mil04} in massless case. We have to take limit $R' \to
\infty$ and then take the derivative with respect $R$ with sign
minus. After integrating by parts we obtain the force in the form
below
\be\label{ForceMilton}
F[\ls,R,R'\to\infty] = -\fr{\partial}{\partial R} E_c^{(1)} = -\fr
1{4\pi R^2} \int_0^\infty \fr{y dy}{(\fr{y}{\ls R} +1) e^y - 1}.
\ee
To compare with Ref. \cite{Mil04} we have to take limit
$\lambda'\to\infty$ in Eq. (2.13) of this paper and change the
notations of parameters $a\to R$ and $\lambda \to \ls R$. As was
noted in Ref. \cite{Mil04} this expression is well defined in the
Dirichlet limit $\ls\to \infty$.

In three dimensional case, $N=3$, one has
\be\label{E3}
E_c^{(3)}[\ls,R,R'] = - \fr{m^4}{12\pi^2} \int_1^\infty dx (x^2 -
1)^{3/2} {\cal Z}_3[mx,\ls,R,R'],
\ee
where
\bnn
&&{\cal Z}_3[k,\ls,R,R'] \\&=& \left\{32 k^5 (R+R') + 8k^3 \ls
\left[-2 + e^{2kR} +e^{2kR'} + 2kR'(e^{2kR}-1) + 2kR
(e^{2kR'}-1)\right] + 4k^2 \ls^2 \left[-2 +
e^{2kR}+ e^{2kR'}\right]\right.\non \\
&+&\left. 2k \ls^3 \left[-2 + e^{2kR}+ e^{2kR'}\right] +
(e^{2kR}-1)(e^{2kR'}-1)\ls^4\right\}\left\{8k^4\left[2k(e^{2k(R+R')}-1)
+ (e^{2kR}-1)(e^{2kR}-1)\ls \right]\right\}^{-1}.\non
\enn
In the limit $R = R' \to \infty$ we obtain the following
expression:
\bd
E_c^{(3)}[\ls,R\to \infty, R'\to \infty] = \fr{m^3}{576 \pi^2}
\left[24\pi - 24 \fr\ls m - 9\pi \fr{\ls^2}{m^2} + 8
\fr{\ls^3}{m^3} - 6(4- \fr{\ls^2}{m^2})^{3/2} \arccos
\fr{\ls}{m}\right],
\ed
which is divergent in the Dirichlet limit $\ls\to\infty$
\bd
E_c^{(3)}[\ls\to \infty,R\to \infty, R'\to \infty] \approx
\fr{\ls^3}{72\pi^2} - \fr{m\ls^2}{64\pi} - \fr{m^2\ls}{32\pi^2} +
\fr{m^3}{24\pi} + \fr{1}{16\pi^2} \left(-\fr{\ls^3}{6} + m^2
\ls\right) \ln\fr\ls m.
\ed

\subsection{A single step function potential}\label{Sec:step}

Let us regularize delta function by step function by relation
\bd
\delta (x) = \lim_{\epsilon\to 0} \left\{ \begin{array}{ll} 0,&
|x| > \epsilon, \\
\fr 1{2\epsilon},& |x| < \epsilon. \end{array}\right.
\ed
and consider the energy for finite value of the regularization
parameter $\epsilon$. In this case we get the following expression
for function $\Psi$ for $R,R' > \epsilon$:
\bd
\Psi_\epsilon^{out}(ik,R,R') = \fr 1k \left[ e^{-kR'} A'_\epsilon
+ e^{kR} A_\epsilon\right],
\ed
where
\bnn
A'_\epsilon &=& e^{kR} \sinh (2\epsilon k_\epsilon)
\fr{k_\epsilon^2 - k^2}{kk_\epsilon} - e^{-kR+2\epsilon k}
\left[\sinh (2\epsilon k_\epsilon) \fr{k_\epsilon^2 +
k^2}{kk_\epsilon} - 2\cosh
(2\epsilon k_\epsilon) \right],\\
A_\epsilon &=& -e^{kR-2k\epsilon} \left[\sinh (2\epsilon
k_\epsilon) \fr{k_\epsilon^2 + k^2}{kk_\epsilon} + 2\cosh
(2\epsilon k_\epsilon) \right] + e^{-kR} \sinh (2\epsilon
k_\epsilon) \fr{k_\epsilon^2 - k^2}{kk_\epsilon},
\enn
and $k_\epsilon^2 = k^2 + \ls/2\epsilon$. In the limit
$\epsilon\to 0$ this function tends to that obtained in last
section and given by Eq. \Ref{PsikRr}:
\be\label{limit}
\lim_{\epsilon\to 0}\Psi_\epsilon^{out}(ik,R,R') = \Psi(ik,R,R').
\ee
The first three heat kernel coefficients are the same as for delta
potential case, the rest coefficients are divergent in the limit
$\epsilon\to 0$:
\bn
B_0^{out} &=& R+R',\ B_{\fr 12}^{out} = -\sqrt{\pi},\non\\
B_n^{out} &=& \fr{2(-1)^n}{n!}\left(\fr\ls 2\right)^n \epsilon^{-n+1}\label{hkkep},\\
B_{n+\fr 12}^{out} &=& (-1)^n \left[\fr{2 \Gamma (n+\fr 12)}{n!} -
\sqrt{\pi}\right] \left(\fr\ls 2\right)^n \epsilon^{-n},\
n=1,2,3,\ldots.\non
\en

To find the function $\Psi$ inside the potential we have to assume
$R,R' < \epsilon$ and consider additional Dirichlet boundaries far
from the potential at points $x=\pm H$. In the end we tend these
boundaries to infinity: $H \gg R,R,\epsilon$. Because we have
three domains confined by Dirichlet boundaries the function $\Psi$
is the product of three functions:
\be
\Psi_\epsilon^{in}(ik,R,R') =
\Psi_\epsilon^1\Psi_\epsilon^2\Psi_\epsilon^3,\label{Psiin}
\ee
where
\bnn
\Psi_\epsilon^1 &=& \fr{e^{-(\epsilon -
R')k_\epsilon}}{4kk_\epsilon} \left\{ e^{k(H-\epsilon)}\left[
e^{-2(\epsilon - R') k_\epsilon} (k_\epsilon - k)
+(k_\epsilon+k)\right] - e^{-k(H-\epsilon)}\left[ e^{-2(\epsilon -
R') k_\epsilon} (k_\epsilon + k) +(k_\epsilon -
k)\right]\right\},\\
\Psi_\epsilon^2 &=& \fr{\sinh (k_\epsilon (R+R'))}{k_\epsilon},\\
\Psi_\epsilon^3 &=& \Psi_\epsilon^1 (R' \Rightarrow R).
\enn
The heat kernel coefficients have the form below
\bn
B_0^{in} &=& 2H,\ B_{\fr 12}^{in} = -3\sqrt{\pi},\non\\
B_n^{in} &=& \fr{2(-1)^n}{n!}\left(\fr\ls 2\right)^n \epsilon^{-n+1}\label{hkkepin},\\
B_{n+\fr 12}^{in} &=& (-1)^n \left[\fr{2 \Gamma (n+\fr 12)}{n!} -
3\sqrt{\pi}\right] \left(\fr\ls 2\right)^n \epsilon^{-n},\
n=1,2,3,\ldots.\non
\en
The difference with above calculated heat kernel coefficients
appears at terms with half-integer indices:
\bd
B_{n+\fr 12}^{in}-B_{n+\fr 12}^{out} = (-1)^{n+1} 2\sqrt{\pi}
\left(\fr\ls 2\right)^n \epsilon^{-n},\ n=0,1,2,\ldots .
\ed

In the framework of zeta-regularization approach \Ref{Emain} the
energy has the following form:
\be\label{ENEp}
E_c^{(N)}[\ls,R,R',\epsilon] = -\fr 1{2(4\pi)^{\fr{N}2}}
\fr{1}{\Gamma (\fr N2 + 1)} \int_m^\infty dk (k^2 - m^2)^{N/2}
{\cal Z}_N[k,\ls,R,R',\epsilon],
\ee
where
\bn
{\cal Z}_N^{out}[k,\ls,R,R',\epsilon] &=& \fr\partial{\partial k}
\ln \Psi_\epsilon^{out}(k,R,R') - (R+R') + \fr 1k\label{RenEp}\\
&-& 2\sum_{l=1}^{\left[\fr{N+1}2\right]} \fr{(-1)^l}{l!}\fr{\Gamma
(l+\fr 12)}{\sqrt{\pi}} \fr{\ls^l}{2^lk^{2l}\epsilon^{l-1}} +
\sum_{l=2}^{\left[\fr{N}2\right]}
\fr{(-1)^l\ls^l}{2^lk^{2l+1}\epsilon^{l}}\left[2 \fr{\Gamma (l+\fr
12)}{\sqrt{\pi}l!} - 1\right]\non, \\
{\cal Z}_N^{in}[k,\ls,R,R',\epsilon] &=& \fr\partial{\partial k}
\ln \Psi_\epsilon^{in}(k,R,R') - 2H + \fr 3k\label{RenEpIn}\\
&-& 2\sum_{l=1}^{\left[\fr{N+1}2\right]} \fr{(-1)^l}{l!}\fr{\Gamma
(l+\fr 12)}{\sqrt{\pi}} \fr{\ls^l}{2^lk^{2l}\epsilon^{l-1}} +
\sum_{l=2}^{\left[\fr{N}2\right]}
\fr{(-1)^l\ls^l}{2^lk^{2l+1}\epsilon^{l}}\left[2 \fr{\Gamma (l+\fr
12)}{\sqrt{\pi}l!} - 3\right]\non .
\en
In the last expression the limit $H\to \infty$ is assumed. Because
of relation \Ref{limit} this expression and the energy are
divergent in the limit $\epsilon\to 0$ starting with dimension
$N=3$.
In the limit of whole space $R=R'=L/2 \to \infty$,
the energy is finite. It is ill defined in the Dirichlet limit,
$\ls \to \infty$, or/and $m\to 0$:
\be\label{Eas1}
E_c^{(N)}[\ls,R,R',\epsilon]_{|m\to 0} \approx \fr{
(-1)^N}{(4\pi)^{\fr{N+1}2}} B_{\fr{N+1}{2}} \ln \fr{m}{\ls}
\ee
where the heat kernel coefficient is given by Eqs. \Ref{hkkep} and
\Ref{hkkepin}.

The energy calculated for this model potential has additional
peculiarity for $R,R'\approx \epsilon$. The energy is divergent in
the limit $R\to \epsilon$ or $R' \to \epsilon$, but it is finite
for $R=\epsilon$ or $R'= \epsilon$. Indeed, let us consider the
asymptotic expansion of function
\bd
\partial_x \ln\Psi_\epsilon^{out}(ik,\epsilon,\epsilon).
\ed
The half-integer coefficients obtained by this expression are
different from \Ref{hkkep} starting from $\fr 32$ coefficient:
\bn
B_0 &=& 2\epsilon,\ B_{\fr 12} = -\sqrt{\pi},\non\\
B_n &=& \fr{2(-1)^n}{n!}\left(\fr\ls 2\right)^n \epsilon^{-n+1}\label{hkkepRtoE},\\
B_{n+\fr 12} &=& (-1)^{n+1} \sqrt{\pi}\left(\fr\ls 2\right)^n
\epsilon^{-n},\ n=1,2,3,\ldots.\non
\en
It is easy understand the origin of this problem. To obtain the
heat kernel coefficients \Ref{hkkep} we assumed that $R,R' >
\epsilon$ and for this reason we turn down all terms as
$e^{-k(R-\epsilon)}$ and $e^{-k(R'-\epsilon)}$ and after this we
set $R=R'=\epsilon$. We may turn down these kind of terms for
arbitrary small, but non-zero, value of $R-\epsilon$ and
$R'-\epsilon$. We obtained the following expression for asymptotic
expansion
\bd
\partial_x \ln\Psi_\epsilon^{out}(ik,R,R')^{as}_{R,R'=\epsilon} =
-\fr 2k + \fr 2{k_\epsilon} - \fr{k}{k_\epsilon^2}+
\fr{2k\epsilon}{k_\epsilon}.
\ed
If we put $R=R'=\epsilon$ at the beginning we obtain another
expression for asymptotic expansion
\bd
\partial_x \ln\Psi_\epsilon^{out}(ik,\epsilon,\epsilon)^{as} =
 - \fr{k}{k_\epsilon^2} + \fr{2k\epsilon}{k_\epsilon}.
\ed
The difference is not zero
\bd
\partial_x \ln\Psi_\epsilon^{out}(ik,R,R')^{as}_{R,R'=\epsilon}
- \partial_x \ln\Psi_\epsilon^{out}(ik,\epsilon,\epsilon)^{as} =
-\fr 2k + \fr 2{k_\epsilon} = -\fr\ls{2\epsilon k^3} + \ldots .
\ed
To reveal this divergence in manifest form let us use another form
of asymptotic expansion. We keep all terms as $e^{-k(R-\epsilon)}$
and $e^{-k(R'-\epsilon)}$ and will consider asymptotic expansion
with these terms as constants:
\bnn
&&\partial_x \ln\Psi_\epsilon^{out}(ik,R,R')^{as}=(R+R') - \fr 1k
- \fr\ls{4\epsilon k^2} \left[2\epsilon -e^{-2 k(R-\epsilon)}
(R-\epsilon) - e^{-2 k(R'-\epsilon)} (R'-\epsilon)\right]\adb\\
&+& \fr\ls{4\epsilon k^3} \left[e^{-2 k(R-\epsilon)}+ e^{-2
k(R'-\epsilon)}\right]\adb\\
&+& \fr{\ls^2}{32\epsilon^2 k^4} \left[6\epsilon + (e^{-4
k(R-\epsilon)} - 2 e^{-2 k(R-\epsilon)}) (R-\epsilon) + (e^{-4
k(R'-\epsilon)} - 2e^{-2 k(R'-\epsilon)}) (R'-\epsilon)\right] +
\ldots
\enn
From this expression we observe that additional terms appears. The
contributions of them to even degree of $k$ (heat kernel
coefficients with integer indices) are insufficient. For $R,R'\not
= \epsilon$ they are exponentially small, but for $R,R' =
\epsilon$ they are zero. The contributions to odd degrees of $k$
(heat kernel coefficients with half-integer indices) are
important. They are exponentially small for $R,R'\not = \epsilon$,
but they are constant for $R,R' = \epsilon$. The first non-trivial
contribution starts from $1/k^3$. Therefore this kind of expansion
reproduces right the expansion for $R,R'\not = \epsilon$ as well
as for $R,R' = \epsilon$.

Let us use this expansion to calculate the zeta-function and the
energy. It allows us to find in manifest form the divergence as
the boundary position $R$ close to boundary of potential
$\epsilon$. In the $3$-dimensional case the divergent contribution
is due to the following term
\bd
-\fr{\ls}{12\pi^2\epsilon}\int_1^\infty (k^2 -
m^2)^{3/2}\fr{e^{-2k(R-\epsilon)}+e^{-2k(R'-\epsilon)}}{k^3} dk.
\ed
For $R-\epsilon \ll 1$ we obtain the following behavior of energy
close to boundary of potential
\be
E_c^{(3)}[\ls,R\approx \epsilon,R',\epsilon] \approx
-\fr{\ls}{12\pi^2\epsilon |R-\epsilon|}.
\ee
The application of this approach to internal function \Ref{Psiin}
gives the same result but with opposite sign.

It is obvious that this divergence connects with model under
consideration. If we adopt another kind of regularization for
delta-function as a consequence of analytic functions we obtain no
divergence except for infinitely close position of plates each to
other as should be the case for Casimir force. For position of
Dirichlet plates exactly on the boundary we obtain the following
finite expression for energy
\be
E_c^{(3)}[\ls,R =\epsilon,R'=\epsilon,\epsilon] =
-\fr{1}{12\pi^2}\int_m^\infty dk (k^2 - m^2)^{3/2} \left[
-2\epsilon + \fr 1k -\fr k{k_\epsilon^2} +
\fr{2k\epsilon}{k_\epsilon} \coth (2\epsilon k_\epsilon) +
\fr\ls{2 k^2} - \fr\ls{2\epsilon k^3} - \fr{3\ls^2}{16\epsilon
k^4}\right].
\ee

\subsection{Three singular potentials $V = \lst\delta (x+R') + \ls
\delta (x) + \lso \delta (x-R)$}\label{Sec:Three}

Let us consider three delta-potentials
\bd
V(x) = \lst\delta (x+R') + \ls \delta (x) + \lso \delta (x-R)
\ed
and the field with Dirichlet boundary condition at surfaces
$x=-l,L$, where $l,L > R',R$.

The $\Psi$ function has the form below
\bnn
&&\Psi(k,R,R',L,l,\ls,\lso,\lst) = \fr{e^{k \left( l + L
\right)}}{k^4} \left\{ -\left( 2 k - \ls \right)\left( 2 k -
{\lso} \right)\left( 2 k - {\lst} \right){e^{-2 k ( l + L)}} \right. \\
&-& \left. {\left( 2 k - \ls \right) {\lso} \left( 2 k - {\lst}
\right)}{e^{-2 k \left( l + R \right) } } + {\ls  {\lso} \left( 2
k - {\lst} \right) } {e^{-2 k \left( l + L - R \right) }  } - {\ls
\left( 2 k + {\lso} \right)\left( 2 k - {\lst}
\right) }{e^{-2 k l}  }\right. \\
&-&\left. {\left( 2 k - \ls \right) \left( 2 k - {\lso}
\right){\lst}}{e^{-2 k \left( L + R' \right) }
 } + {\ls  \left( 2 k - {\lso} \right)  {\lst}}{e^{-2 k
\left( l + L - R' \right) }} + {\ls  {\lso} {\lst}} {e^{-2
k \left( L + R' - R \right) }}\right.\\
&+&\left.{\ls  {\lso} {\lst}}{e^{-2 k \left( l - R' + R \right) }
} - {\left(2 k - \ls  \right)  {\lso} {\lst}} {e^{-2 k \left( R' +
R \right) }  } + {\left( 2 k + \ls \right) {\lso}{\lst}}{e^{-2 k
\left(l + L - R' - R \right) } } - {\ls \left( 2 k + {\lso}
\right){\lst}}{e^{-2 k R'}  }\right.\\
&-&\left. {\left( 2 k + \ls \right)\left( 2 k +
{\lso}\right){\lst}}{e^{-2 k \left(l - R' \right) } }- {\ls \left(
2 k - {\lso} \right)\left( 2 k + {\lst} \right) }{e^{-2 k L}  } -
{\ls  {\lso} \left( 2 k + {\lst} \right) }{e^{-2 k R}  }\right.\\
&-& \left.{\left( 2 k + \ls \right) {\lso} \left( 2 k + {\lst}
\right) }{e^{-2 k \left(L - R \right) } } +  {\left( 2 k + \ls
\right) \left( 2 k + {\lso} \right) \left( 2 k + {\lst} \right) }
\right\}.
\enn
For $\lso=\lst=0$ this function coincides to that considered in
the first section. The following relation is valid:
\bd
\Psi(k,R,R',L,l,2\ls,\lso=0,\lst=0) =
\Psi(k,R=0,R'=0,L,l,\ls=0,\ls,\ls)
\ed
as should be the case. But the energy does not obey this kind of
relation. Indeed, let us consider the case of two potentials and
put $\ls =0$. One has
\bn
&&\Psi(k,R,R',L,l,\ls=0,\lso,\lst) = \frac{2e^{2 k L}}{k^3}
\left\{ -{\left( 2 k - \lso \right)\left( 2 k - \lst \right)
}{e^{-4 k L}} - {\lso \left( 2 k - \lst \right) }{e^{-2 k \left( L
+ R \right) }}\right.\non\\
&-&\left. {\left( 2 k - \lso \right) \lst}{e^{-2 k \left( L + R'
\right) }} - {\lso \lst}{e^{-2 k \left( R' + R \right) }} +
{\lso \lst}{e^{-2 k \left( 2 L - R' - R \right) }}\right.\label{PsiTwo}\\
&-&\left. {\left( 2 k + \lso \right)\lst}{e^{-2 k \left(L - R'
\right) }} - {\lso \left( 2 k + \lst \right)}{e^{-2 k \left( L - R
\right) }}  + \left( 2 k + \lso \right)  \left( 2 k + \lst
\right)\right\}.\non
\en
In this case
\bnn
{\cal Z}_N[k,0,\lso,\lst,R,R'] &=& \fr\partial{\partial k} \ln
\Psi(k,R,R',L,l,0,\lso,\lst) - (L+l) + \fr 1k - \sum_{l=1}^N
(-1)^l \fr{\lso^l+\lst^l}{2^lk^{l+1}},\\
{\cal Z}_N[k,\ls,0,0,R,R'] &=& \fr\partial{\partial k} \ln
\Psi(k,R,R',L,l,\ls,0,0) - (L+l) + \fr 1k - \sum_{l=1}^N (-1)^l
\fr{\ls^l}{2^lk^{l+1}}.
\enn
The difference has the following form:
\bd
\triangle{\cal Z} = - \sum_{l=2}^N (-1)^l \fr{2\ls^l -
(2\ls)^l}{2^lk^{l+1}},
\ed
and it is non-zero starting from dimension $N=3$. This fact
originates from renormalization procedure.

In the limit of whole space, $L=l\to\infty$, we get from
\Ref{PsiTwo}
\bn\label{PsiTwoLim}
&&\Psi(k,R,R',L,l,\ls=0,\lso,\lst) = \frac{2e^{2 k L}}{k^3}
\left\{- {\lso \lst}{e^{-2 k \left( R' + R \right) }} + \left( 2 k
+ \lso \right)  \left( 2 k + \lst \right)\right\}.
\en
If we then put potentials to infinity $R=R'\to\infty$ we obtain
non-zero result
\bnn
{\cal Z}_N[k,0,\lso,\lst,R,R'] &=& -\fr{\lso}{k(2k+\lso)} -
\fr{\lst}{k(2k+\lst)} - \sum_{l=1}^N (-1)^l
\fr{\lso^l+\lst^l}{2^lk^{l+1}}\\
&=&- \sum_{l=N+1}^\infty (-1)^l \fr{\lso^l+\lst^l}{2^lk^{l+1}}.
\enn
This is in contradiction with expected result that the energy has
to be zero.
We observe that in this limit the energy is the sum of energy of
two single plates. Therefore we conclude that this energy is
connected with potential itself and it has the additivity property
provided the infinite distances between potentials (without
interaction). For $J$ plates in the limit of infinite distances we
get
\bd
{\cal Z}_N = -\sum_{i=1}^J\fr{\lambda_{si}}{k(2k+\lambda_{si})}
-\sum_{i=1}^J \sum_{l=1}^N (-1)^l \fr{\lambda_{si}^l}{2^lk^{l+1}}.
\ed

Therefore, with each $j$-th singular potential $\lambda_{sj}\delta
(x-R_j)$ we may connect the "surface"\/ energy by relation
\be\label{ENSur}
E_{sj}^{(N)} = \fr{m^{N+1}}{2(4\pi)^{\fr{N}2}} \fr{1}{\Gamma (\fr
N2 + 1)} \int_1^\infty dx (x^2 - 1)^{N/2}\left[
\fr{\lambda_{sj}}{mx(2mx+\lambda_{sj})} + \sum_{l=1}^N
\fr{(-1)^l\lambda_{sj}^l}{2^lm^{l+1}x^{l+1}}\right].
\ee
This energy is ill defined in the limits $m\to 0$ or/and
$\lambda_{sj} \to \infty$.

\section{Renormalization procedure}\label{SecRenormalization}

Let us summarize useful information from above section. First of
all, we should like to note that the divergence of above energies
is connected with renormalization procedure only. The regularized
expression for energy \Ref{Edef}, \Ref{zetaN} is well defined in
the limit $\ls\to \infty$ or $\epsilon \to 0$. The renormalization
procedure brings a problem. The point is that for renormalization
we subtract finite number of terms (see \Ref{Ren} or \Ref{RenEp})
of series expansion over small value $\ls/k \ll 1$ or
$\ls/\epsilon k^2 \ll 1$. After renormalization we use expressions
obtained for $\ls \to \infty$ or $\epsilon\to 0$. Obviously that
the main divergence originate from the last term of truncated
series. Indeed, let us consider the asymptotic expansion over $k$
the integrand
\be\label{PsiSeries}
\fr\partial{\partial k} \ln \Psi(k,R,R')^{as} = R+R' - \fr 1k -
\fr{\ls}{k(2k+\ls)} + O(e^{-kR},e^{-kR'}).
\ee
The last term is finite in the limit $\ls\to \infty$ and the
difference
\be\label{difference}
\fr\partial{\partial k} \ln \Psi(k,R,R') - \fr\partial{\partial k}
\ln \Psi(k,R,R')^{as}
\ee
is finite in this limit, too. But if we use truncated series
expansion
\bd
\fr\partial{\partial k} \ln \Psi(k,R,R')^{as}= R+R' - \fr 1k +
\sum_{l=1}^{N} (-1)^l \fr{\ls^l}{2^lk^{l+1}}
\ed
the difference \Ref{difference} is infinite in the Dirichlet or
massless limit. The main divergence comes from the last term which
gives the contribution to heat kernel coefficient $B_{\fr{N+1}2}$
which in turn gives the conformal anomaly.

Second, with each potential connects the inner quantity given by
Eq. \Ref{ENSur} which we called "surface"\/ energy. It is defined
as energy of single delta-potential provided the boundaries are in
infinity. This quantity possesses an additive property: the
"surface"\/ energy of set delta-potentials is the sum of
"surface"\/ energies of each delta-potential provided the infinite
distance between potentials. We note also that this quantity is
connected with potential itself. In self-consistent theory which
takes into account gravitational field of brane \cite{Vil81}, this
energy will give some contribution to gravitational field of
brane.

For this reason let us consider another renormalization procedure.
Instead of extracting terms which will survive in the limit $m\to
\infty$ we subtract all terms in asymptotic expansion of zeta
function. It is easy to see that this procedure corresponds to
subtracting "surface"\/ energy of delta potentials, which is
defined for $R,R'\to\infty$. It seems reasonable that the Casimir
energy is zero if all potentials are in infinity. Indeed, all
terms of asymptotic expansion in \Ref{PsiSeries} may be obtained
by taking the limits $R,R' \to \infty$:
\bd
\fr\partial{\partial k} \ln \Psi(k,R,R')_{R,R'\to \infty} = R+R' -
\fr 1k - \fr{\ls}{k(2k+\ls)} + O(e^{-kR},e^{-kR'}).
\ed
It is easy to see that the condition \Ref{condition} is still
valid for this kind of renormalization. In this case the energy
for single delta-potential considered in Sec. \ref{Sec:Single} has
the form \Ref{EN}
\be\label{ENew}
\widetilde{E}_c^{(N)}[\ls,R,R'] = -\fr 1{2(4\pi)^{\fr{N}2}}
\fr{1}{\Gamma (\fr N2 + 1)} \int_m^\infty dk (k^2 - m^2)^{N/2}
\widetilde{{\cal Z}}_N[k,\ls,R,R'],
\ee
where function
\be\label{ZOther}
\widetilde{{\cal Z}}_N[k,\ls,R,R'] = \fr{2\left(4 k^2 (R+R') + \ls
\left[-2 + e^{2kR}(1+2kR') + e^{2kR'}(1+2kR)\right] +
\ls^2\left[(e^{2kR'}-1)R + (e^{2kR}-1)R'\right]\right)}
{(2k+\ls)\left[2k(e^{2k(R+R')}-1) + (e^{2kR}-1)(e^{2kR}-1)\ls
\right]}
\ee
does not depend on the dimension of space. It is easy to see that
the difference of these energies is that divergent expression
\Ref{Ediv} which we called "surface"\/ energy \Ref{ENSur}:
\bd
E_c^{(N)}[\ls,R,R'] - \widetilde{E}_c^{(N)}[\ls,R,R'] =
E_c^{(N)}[\ls,R\to\infty,R'\to\infty] = E_{s}^{(N)}.
\ed

The renormalization procedure suggested by Lukosz \cite{Luk71}
(see also \cite{DowKen78}) gives the same result. Indeed,
according with this approach we have to include our system to
large box, for example with $x=\pm H$, where $H> R,\ R'$. The
space is divided into three domains: ${\cal M}_1: \{x\in
[-H,-R']\}$, ${\cal M}_2: \{x\in [-R',R]\}$ and ${\cal M}_3:
\{x\in [R,H]\}$. The renormalized energy is given by equation
\bd
\triangle E_c^{(N)} = E_c^{(N)}({\cal M}_1)+E_c^{(N)}({\cal M}_2)
+ E_c^{(N)}({\cal M}_3) - E_c^{(N)}({\cal M}),
\ed
where the last term is calculated for whole space but without
internal boundaries. In the end we have to tend $H\to\infty$. It
is very easy to apply this formula in our case. Because in domains
${\cal M}_1$ and ${\cal M}_2$ there is no potential we may use our
formulas with $\ls = 0$. Therefore we obtain
\bnn
\triangle E_c^{(N)} &=& E^{(N)}_c [\ls = 0, -R',H] + E^{(N)}_c
[\ls , R,R'] + E^{(N)}_c [\ls = 0, H,-R] - E^{(N)}_c [\ls,H,H]\\
&=& -\fr{1}{2(4\pi)^{\fr{N}2}} \fr{1}{\Gamma (\fr N2 + 1)}
\int_m^\infty dx (k^2 -m^2)^{3/2} \fr\partial{\partial k} \ln
\fr{k^2\Psi(ik,-R',H)_{\ls=0} \Psi(ik,R,R') \Psi(ik,H,-R)_{\ls =
0}}{\Psi(ik,H,H)}.
\enn
In the limit $H\to\infty$ we obtain the expression for the energy
which coincides exactly  with that obtained above and given by Eq.
\Ref{ENew}.

Let us consider some special cases of energy obtained. In massless
limit the energy is finite for any dimensions. In the Dirichlet
limit $\ls\to\infty$,
\bd
\widetilde{{\cal Z}}_N[k,\ls\to\infty,R,R'] = \fr{2R}{e^{2kR}-1} +
\fr{2R'}{e^{2kR'}-1},
\ed
the energy, as expected, is finite and it is the sum of two
Casimir energies for domains $(-R',0)$ and $(0,R)$:
\bd
\widetilde{E}_c^{(N)}[\ls\to\infty,R,R'] =
\widetilde{E}_c^{(N)}[\ls,0,R']+\widetilde{E}_c^{(N)}[\ls,R,0],
\ed
as should be the case. For $\ls = 0$
\bd
\widetilde{{\cal Z}}_N[k,\ls=0,R,R']= \fr{2(R+R')}{e^{2k(R+R')}-1}
\ed
which gives the Casimir energy for two plates in points $-R'$ and
$R$. Therefore, this energy has satisfactory behavior for all
reasonable cases. In one-dimensional case $N=1$ the force acting
for one plate $(R'\to \infty)$ due to singular potential coincides
with that obtained by Milton in Ref. \cite{Mil04} in massless case
and it is given by Eq. \Ref{ForceMilton}. This fact is evident
because the difference of two energies is constant  which does not
depend on the distance to boundary.

\begin{figure}[h]
\begin{center}
\epsfxsize=9truecm\epsfbox{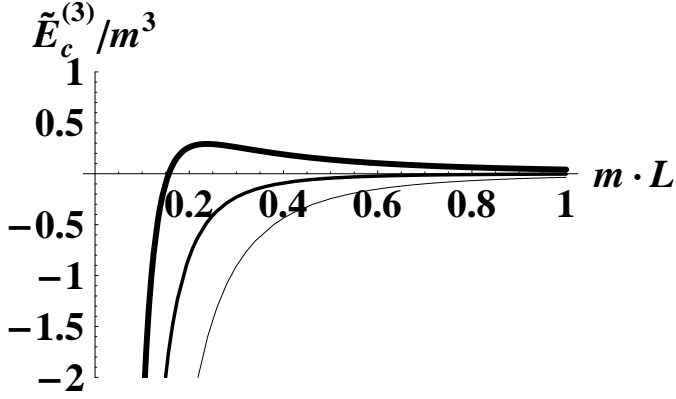}
\end{center} \caption{The Casimir energy for two plates in
positions $x=\pm L$ with singular potential $V=\ls\delta (x)$
between them for $\ls=-1.9 m$ (thick curve), $\ls=0$ (middle
curve), and $\ls=4m$ (thin curve).} \label{gr}
\end{figure}

The dependence of the Casimir energy \Ref{ENew} for three
dimensional case is plotted in Fig. \ref{gr} for $R=R'=L/2$ as a
function of $Lm$ for different values of $\ls$ and the fixed mass
$m$.
We observe that the energy is negative for $\ls \ge 0$. It is easy
to see this statement from Eq. \Ref{ZOther}: in this case all
terms of ${\cal Z}$ are positive. The slope of curves is negative
for all position of plates. It means that two plates are attracted
as in usual Casimir case. For negative $\ls$ a maximum appears at
position $L_m$ and the energy in the maximum growths with
increasing $|\ls|$. For distances smaller then $L_m$ the plates
are attracted but for $L>L_m$ the force is repulsive. The
repulsive character of Casimir force due to delta potential was
observed in Ref. \cite{Sca99}. Because of the present theory is
valid for $\ls/m
> -2$ we obtain from Eq. \Ref{BoundIneq} that for $Lm>2$ the
boundary state appears. The energy of this state $E = \sqrt{m^2 -
k^2}$ is found from following equation
\bd
\tanh \fr{kL}2 = - \fr{2k}\ls.
\ed
It is easy to show from Eq. \Ref{ZOther} that the energy is
positive for large distance between plates $mL\to \infty$ and for
$0>\ls/m>-2$. Therefore for large distances between plates the
Casimir force is repulsive and boundary state appears localized at
the position of potential. There is a close analogy of this point
with widely discussed now "surface states"\/ \cite{IntLam05}. It
was shown the repulsive character of "surface states" which may be
regarded as boundary states \cite{IntLam05}.

In the case of step function potential the Lukosz regularization
or renormalization procedure suggested above give the following
expression for energy ($R,R'> \epsilon$)
\be\label{Eepnew}
\widetilde{E}^{(3)}_c(\ls,R,R',\epsilon) = -
 \fr{1}{12\pi^2} \int_m^\infty dk (k^2 -m^2)^{3/2}
\fr\partial{\partial k} \ln
\fr{k^2\Psi_\epsilon^{out}(ik,-R',H)_{\ls=0}
\Psi_\epsilon^{out}(ik,R,R') \Psi_\epsilon^{out}(ik,H,-R)_{\ls =
0}}{\Psi_\epsilon^{out}(ik,H,H)|_{H\to \infty}}.
\ee
For the case of position of plates inside the potential, $R,R' <
\epsilon$ we put the system in large box with size $2H$. We have
three domains: $x\in (-H,-R'),\ x\in (-R',R),\ x\in (R,H)$. For
renormalization we have to subtract the energy without internal
boundaries at $R$ and $-R'$. This energy is defined by function
$\Psi^{out}$ because the boundaries at $\pm H$ are outside the
potential. Therefore one has:
\be\label{EepnewIn}
\widetilde{E}^{(3)}_c(\ls,R,R',\epsilon) = -
 \fr{1}{12\pi^2} \int_m^\infty dk (k^2 -m^2)^{3/2} \left[
\left.\fr\partial{\partial k} \ln \fr{k^2
\Psi_\epsilon^{in}(ik,R,R')}{\Psi_\epsilon^{out}(ik,H,H)}\right|_{H\to
\infty} - \fr\ls{\epsilon k^3}\right].
\ee

Due to relation \Ref{limit} it is obvious that in the limit
$\epsilon\to 0$ the energy \Ref{Eepnew} is finite and it coincides
with energy calculated above for $\delta$ potential
\bd
\widetilde{E}^{(3)}_c(\ls,R,R',\epsilon\to 0) =
\widetilde{E}^{(3)}_c(\ls,R,R').
\ed
In the limit $\ls\to \infty$ one has
\be
E^{(3)}_c(\ls\to\infty,R,R',\epsilon) = -
 \fr{1}{12\pi^2} \int_m^\infty dk (k^2 -m^2)^{3/2}
\left[\fr{2(R-\epsilon)}{e^{2k(R-\epsilon)}-1} +
\fr{2(R'-\epsilon)}{e^{2k(R'-\epsilon)}-1}\right]
\ee
which is exact the sum of two Casimir energies for domains $x\in
[-R',-\epsilon]$ and $x\in [\epsilon,R]$ as should be the case.
For zero potential $\ls =0$ the expression \Ref{Eepnew} reproduce
the Casimir energy for two boundaries at points $x=-R',R$.

\begin{figure}[h]
\begin{center}
\epsfxsize=8.5truecm\epsfbox{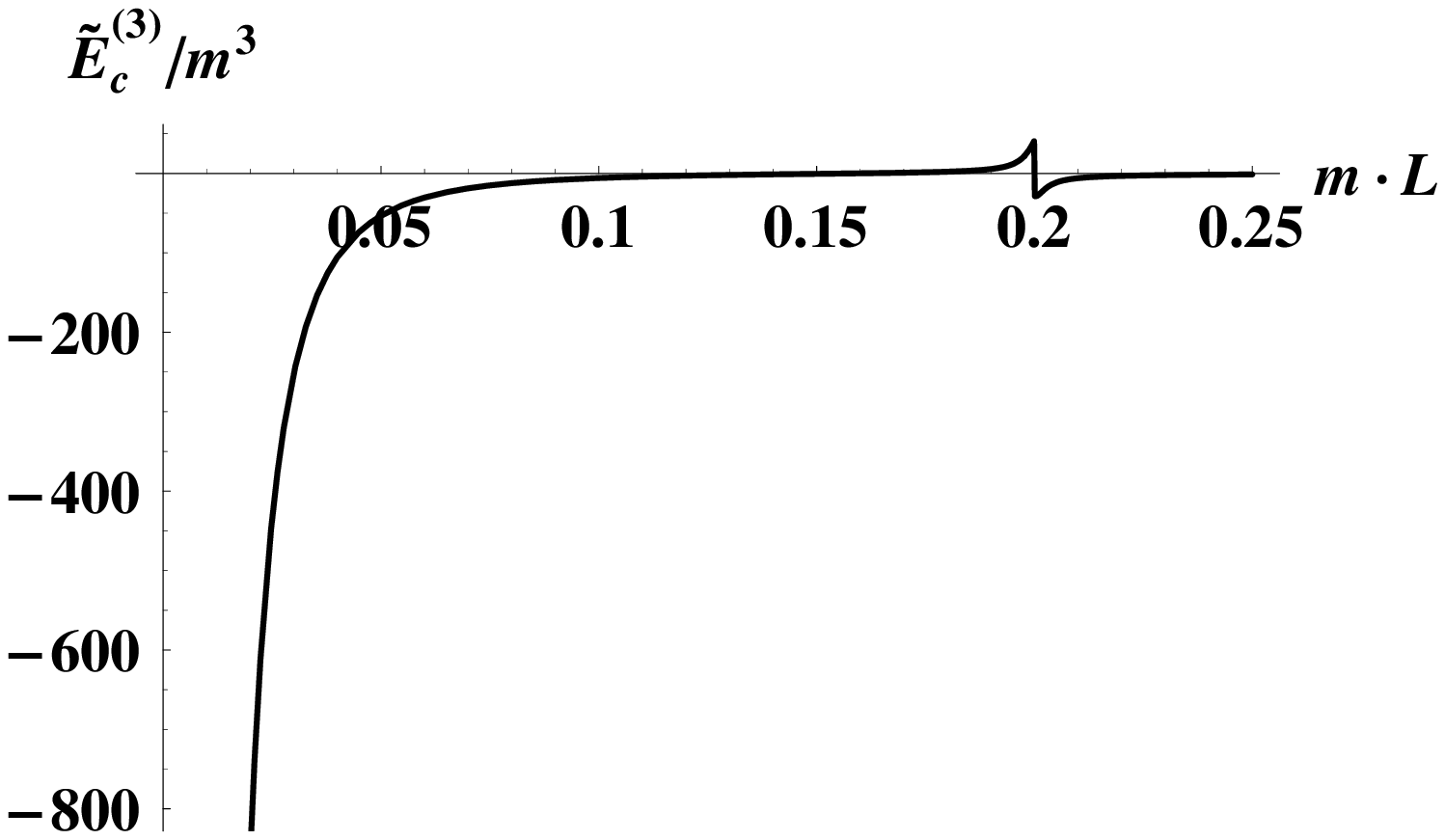}%
\epsfxsize=8.5truecm\epsfbox{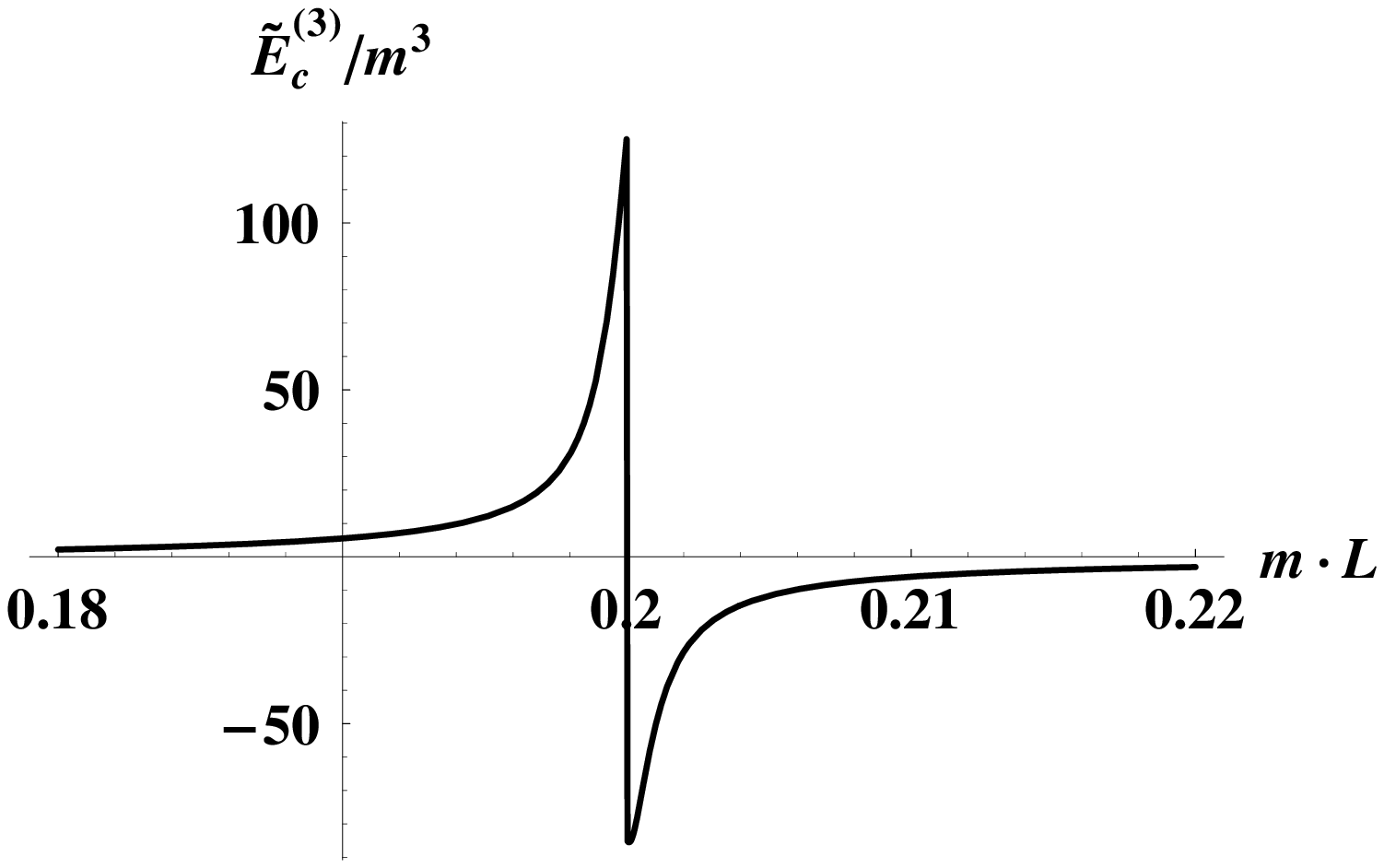}
\end{center} \caption{The Casimir energy for $3$-dimensional case and
for $R=R'=L/2, \ls =1, \epsilon = 0.1$. The boundary of potential
is at point $Lm = 0.2$. There is a divergence in this point which
is plotted in detail in the right figure.} \label{ep12}
\end{figure}

The numerical calculations of Casimir energy is reproduced in Fig.
\ref{ep12} for $R=R'=L/2,\ls =1,\epsilon = 0.1$. We observe the
divergence of energy at the boundary of potential $L=2\epsilon =
0.2$ which was noted in Sec. \ref{Sec:step}. For all position of
Dirichlet plates the Casimir force is attractive. In the right
figure we show details of the energy close to this boundary. The
smaller $\epsilon$ the closer the position of this divergence to
origin and the closer external part to the case calculated for
delta-function case. For small $\epsilon$ and negative $\ls$ the
Fig. \ref{gr} is reproduced.

In the case of three delta functions in framework of this
procedure we obtain the following expression for the function
$\widetilde{{\cal Z}}$:
\bd
\widetilde{{\cal Z}}_N = \Psi(k,R,R',L,l,\ls,\lso,\lst) \fr{e^{- k
\left( l + L \right)}k^4}{\left( 2 k + \ls \right) \left( 2 k +
{\lso} \right) \left( 2 k + {\lst} \right) }.
\ed
The same expression may be obtained by Lukosz renormalization
procedure. We have to use this approach for each solitary
potential provided that others are in the infinity. This
expression is well defined in all limits we need. For example, in
Dirichlet limit $\ls, \lso, \lst \to \infty$ the energy is the sum
of fore Casimir energies for plates in points $x=-l,-R',0,R,L$.

\section{Effective action point of view}\label{Sec:Eff}

Let's consider the same problem from the point of view of
effective action. One has the scalar field $\varphi$ living in
$(3+1)$ dimensional space-time with singular potential
concentrated at the surface $\Sigma$. To make clear regularization
procedure we consider a nonlinear theory with the following action
\be\label{Action}
{\cal S} = \fr 12\int_M \left((\nabla\varphi)^2 + m^2 \varphi^2 +
\fr\lambda 6 \varphi^4 + \Lambda\right)d^4x + \fr 12\int_{\Sigma}
\left(\Lambda_s + \ls \varphi^2\right)d^3x.
\ee
Here the $\Lambda$ is cosmological constant and $\Lambda_s$ is
surface tension. As was noted in Refs. \cite{GeoGraHai01,BorVas04}
we have to include the surface terms at the beginning for correct
renormalization procedure. In fact, the action \Ref{Action}
describes a brane with co-dimension one in four dimensional
space-time with tension $\Lambda_s$ and mass of brane $\ls$. The
renormalization group of brane with co-dimension more then unit
was considered  in Ref. \cite{GolWis01}. Similar problem for free
scalar field without self-action was considered in Ref.
\cite{Tom05}.

To calculate the energy of quantum fluctuation we consider the
field at finite temperature $T = 1/\beta$ and in the end of
calculation we take the limit of zero temperature
$\beta\to\infty$. For this reason we proceed to the Euclidean
regime $t \to -i\tau$ with $\tau \in [0,\beta]$. Let us consider
the Casimir problem -- two Dirichlet plates at points $x=R$ and
$x=-R'$. The singular potential is concentrated at point $x=0$.

We divide the field on classical part and quantum fluctuations:
$\varphi = \varphi_{cl} + \phi$. In one loop approximation and
framework of zeta regularization approach we have the following
expression for action
\bd
{\cal S} ={\cal S}_{cl} + {\cal S}_q= \fr 12\int_M
\left((\nabla\varphi_{cl})^2 + m^2 \varphi_{cl}^2 + \fr\lambda 6
\varphi_{cl}^4 + \Lambda\right) + \fr 12\int_{\Sigma}
\left(\Lambda_s + \ls \varphi_{cl}^2\right) - \fr 12
\widetilde{\mu}^{2s} \Gamma (s) \zeta(s,D_4),
\ed
where
\be
D_4 = -\triangle_4 + m^2 + \lambda \varphi_{cl}^2 + \ls\delta(x).
\ee
In the limit $s\to 0$
\bd
{\cal S}_q = -\fr 12 \left(\fr 1s + \ln \mu^2\right) \zeta(0,D) -
\fr 12 \zeta'(0,D),
\ed
where $\mu^2 = \widetilde{\mu}^2e^{-\gamma}$.

Due to the relation
\be\label{zeta4}
\zeta(s,D_4) = \fr 1{4\pi} \fr{\zeta(s-1,D_2)}{s-1}
\ee
we consider two dimensional eigenvalue problem
\bd
\left[-\fr{\partial }{\partial x} - \fr{\partial }{\partial \tau}
+ M^2 + \ls\delta(x)\right]\phi = \omega^2 \phi
\ed
with $M^2 = m^2 + \ls \varphi_{cl}^2$. Taking into account the
periodicity over $\tau$ we obtain
\bd
\left[-\fr{\partial }{\partial x} + \left(\fr{2\pi
n}{\beta}\right)^2 + M^2 - \omega^2 + \ls\delta(x)\right]\phi = 0,
\ed
where $n=0,\pm 1,\pm 2, \ldots$. Let us denote $\left(\fr{2\pi
n}{\beta}\right)^2 + M^2 - \omega^2 = p^2$ and consider the
solution of above equation in imaginary axis $p=ik$. In this case
the equation reads
\be
\left[-\fr{\partial^2 }{\partial x^2}  + k^2  +
\ls\delta(x)\right]\phi = 0.
\ee

This equation has already considered in last section. Using the
results we arrive at the following expression for zeta function
\be
\zeta(s,D_2) = \fr{\sin (\pi s)}{\pi} \sum_{n=-\infty}^{\infty}
\int_{M_n}^\infty dk (k^2 - M_n^2)^{-s} \fr\partial{\partial k}
\ln \Psi(k,R,R'),\label{main}
\ee
where $M_n^2 = M^2 + (2\pi n/\beta)^2$.

Taking into account this expression we obtain heat kernel
coefficients
\bnn
B_0 &=& \Omega,\ B_{\fr 12} = -\sqrt{\pi}\Sigma,\\
B_{n} &=& - \left(\fr\ls 2\right)^{2n-1}
\fr{\sqrt{\pi}\Sigma}{\Gamma
(n+\fr 12)},\\
B_{n+\fr 12} &=& \left(\fr\ls 2\right)^{2n}
\fr{\sqrt{\pi}\Sigma}{n!},\ n=1,2,3,\ldots .,
\enn
where $\Omega = (R+R')\beta$ is the volume and $\Sigma = \beta$ is
the square of single boundary surface. These heat kernel
coefficients may be obtained by multiplication to $\Sigma$ the
coefficients obtained for delta potentials \Ref{hkcdelta}. We
observe that the zeroth coefficient is volume of the system and
all others coefficients are proportional to square of boundary. To
note this moment it is better to define the density $b_n$ of heat
kernel coefficients by relations $B_0 = b_0\Omega$ and $B_n =
b_n\Sigma,\ n>0$. In calculations of heat coefficients the
asymptotic relation \cite{Eli94}
\bd
\sum_{n=-\infty}^{+\infty} M_n^{-l-2s}{}_{|\beta\to\infty} = \fr
1{2\sqrt{\pi}} \fr{\Gamma (s+\fr l2 - \fr 12)}{\Gamma (s+\fr l2)}
m^{1-l-2s}\beta
\ed
was used.

To obtain the effective action we have to calculate
$\zeta(-1,D_2)$ and $\zeta'(-1,D_2)$. For this aim we will use the
formula \Ref{main}. We subtract from and add to integrand the
first five terms of asymptotic expansion of it and represent the
zeta function in the form
\bn
\zeta(s,D_2)&=&\zeta_a(s,D_2)+\zeta_b(s,D_2)\label{zetaexp}\\
&=& \fr{\sin (\pi s)}{\pi} \sum_{n=-\infty}^{\infty}
\int_{M_n}^\infty dk (k^2 - M_n^2)^{-s} {\cal Z}_3[k,\ls,R,R']\non\\
&+& \fr{1}{4\pi} \fr 1{\Gamma(s)} \sum_{n=0}^4 B_{\fr n2} m^{-2 s
+ 2 -n}\Gamma(s - 1 + \fr n2).\non
\en

Using this equation we calculate the zeta function at point $-1$:
\bnn
\zeta_a(-1,D_2) &=&0,\\
\zeta_b(-1,D_2) &=& \fr 1{4\pi} \left\{-\fr 12 m^4\Omega b_0 +
\Sigma [ m^2 b_1 - b_2]\right\}\non
\enn
and its derivative in this point
\bnn
\zeta'_a (-1,D_2) &=& - \sum_{n=-\infty}^{\infty}
\int_{M_n}^\infty dk (k^2 - M_n^2) {\cal Z}_3[k,\ls,R,R'],\\
\zeta'_b (-1,D_2) &=& \fr{\ln m^2}{4\pi} \left\{\fr 12 m^4\Omega
b_0 - \Sigma [ m^2 b_1 - b_2]\right\} - \fr 1{4\pi} \left\{\fr 14
m^4\Omega b_0 + \Sigma\left[\fr{4\sqrt{\pi}}3 m^3 b_{\fr 12} -
2\sqrt{\pi} m b_{\fr 32} - b_2\right]\right\}\non .
\enn
To calculate $\zeta'_a (-1,D_2)$ we use the Abel-Plana formula
\bd
\sum_{n=1}^\infty f[n]= \int_0^\infty f[x] dx - \fr 12 f[0] + i
\int_0^\infty \fr{f[iy+\epsilon]- f[-iy+\epsilon]}{e^{2\pi y} - 1}
dy.
\ed
In the limit $\beta\to\infty$ we get
\bd
\zeta'_a (-1,D_2) = -\fr{2m^4\beta}{3\pi} \int_1^\infty dx (x^2 -
1)^{3/2} {\cal Z}_3[mx,\ls,R,R'] + O(e^{-m\beta}).
\ed
Then, using Eq. \Ref{zeta4} we express four dimensional zeta
function in therms of two dimensional:
\bnn
\zeta(0,D_4) = - \fr 1{4\pi} \zeta(-1,D_2),\\
\zeta'(0,D_4) = - \fr 1{4\pi} \zeta'(-1,D_2)- \fr 1{4\pi}
\zeta(-1,D_2).
\enn

Taking into account these equations and putting all terms together
we obtain the following expression for effective action
\be\label{S}
{\cal S} = \Omega {\cal S}^{eff}_{\Omega} + \Sigma {\cal
S}^{eff}_{\Sigma} + \beta E_c^{(3)} + O(e^{-m\beta}),
\ee
where
\bnn
{\cal S}^{eff}_{\Omega} &=& b_0\left(-\fr{M^4}{64\pi^2 s} -
\fr{3M^4}{128\pi^2} + \fr{M^4}{64\pi^2} \ln \fr{M^2}{\mu^2}\right)
+ \fr 12 m^2 \varphi_{cl}^2 + \fr\lambda
{12} \varphi_{cl}^4 + \fr 12 \Lambda,\\
{\cal S}^{eff}_{\Sigma} &=& \fr{M^2b_1 - b_2}{32\pi^2s}  -
\fr{M^3b_{1/2}}{24\pi^{3/2}} + \fr{M^2b_{1}}{32\pi^{2}} +
\fr{Mb_{3/2}}{16\pi^{3/2}} - \fr{M^2b_1 - b_2}{32\pi^2}\ln
\fr{M^2}{\mu^2} + \fr 12 \Lambda_s + \fr 12 \ls \varphi_{cl}^2,
\enn
and $E_c^{(3)}$ is given by Eq. \Ref{E3}, and $M^2 = m^2 + \lambda
\varphi_{cl}^2$. The last term in Eq. \Ref{S} can not be regarded
as volume or surface contribution.

From above expression for effective action we observe that one has
to renormalize not merely parameters of volume part of action but
also parameters of surface contribution, too. To renormalize
effective action we use minimal subtraction procedure and make the
following shift of constants
\be\label{RenConst}
m^2 \to m^2 + \fr{\lambda m^2}{16\pi^2 s},\ \lambda \to \lambda +
\fr{3\lambda^2}{16\pi^2 s}, \ \Lambda \to \Lambda +
\fr{m^4}{32\pi^2 s},\ \ls \to \ls - \fr{\lambda b_1}{16\pi^2 s},\
\Lambda_s \to \Lambda_s - \fr{m^2 b_1 - b_2}{16\pi^2 s}.
\ee
The renormalization of the surface parameters depends on the
details of potential (brane) which are encoded in heat kernel
coefficients. The same renormalization condition was used in Ref.
\cite{BorVas04}. We may realize our model as a model of scalar
field in curved space-time but with singular scalar curvature as,
for example, in Ref. \cite{KhuSus02} for wormhole space-time. In
this case the parameter $\Lambda_s$ is proportional to radius of
wormhole's throat $a$ and $\ls$ plays the role of non-conformal
coupling $\ls \sim a\xi$ and it has to be renormalized, too, as
was noted in Ref. \cite{Odi91}. Therefore the effective action
takes the following form
\bnn
{\cal S}^{eff}_{\Omega} &=& - \fr{3M^4}{128\pi^2} +
\fr{M^4}{64\pi^2} \ln \fr{M^2}{\mu^2} + \fr 12 m^2 \varphi_{cl}^2
+ \fr\lambda {12} \varphi_{cl}^4 + \fr 12 \Lambda,\\
{\cal S}^{eff}_{\Sigma} &=& - \fr{M^3b_{1/2}}{24\pi^{3/2}} +
\fr{M^2b_{1}}{32\pi^{2}} + \fr{Mb_{3/2}}{16\pi^{3/2}} - \fr{M^2b_1
- b_2}{32\pi^2}\ln \fr{M^2}{\mu^2} + \fr 12 \Lambda_s + \fr 12 \ls
\varphi_{cl}^2.
\enn
To fix the values of parameters of model both volume and surface
parts we have the only free parameter $\mu$. To remove the
renormalization unambiguous in effective potential we may use the
standard relation for effective potential
\be
{\cal S}^{eff}_{\Omega}(0)'' = m^2,
\ee
by using which we obtain that $\ln m^2/\mu^2 = 1$. This value of
$\mu$ we exploit in surface part of effective action. Therefore
there is no way to fix surface parameters of the theory and
ambiguous in renormalization procedure in surface term appears
which was noted in Ref. \cite{BorVas04}. The similar problem
appears in Ref. \cite{Fur94} where effective action was calculated
in space-time of cosmic string. We arrive at the following
expression for action for zero value of classical field
\bnn
{\cal S}^{eff}_{\Omega}(0) &=& - \fr{m^4}{128\pi^2} + \fr 12 \Lambda,\\
{\cal S}^{eff}_{\Sigma}(0) &=& - \fr{m^3b_{1/2}}{24\pi^{3/2}} +
\fr{mb_{3/2}}{16\pi^{3/2}} + \fr{b_2}{32\pi^2}  + \fr 12
\Lambda_s\\
&=& \fr{m^3}{24\pi} + \fr{m\ls^2}{64\pi} - \fr{\ls^3}{192\pi^2} +
\fr 12 \Lambda_s,
\enn
where the last term is defined ambiguously up to finite terms. We
may make consequent arbitrary finite renormalization. To avoid
this ambiguous let us, in spirit of last section, subtract from
effective action all terms which will survive in the limit of
infinite distance between plates, that is the "surface"\/ energy.
It corresponds to reasonable condition that the energy in empty
space is zero. In this case the energy of the system is
\be\label{energy}
E = \fr{\partial}{\partial \beta}\left[ {\cal S}(0)- \lim_{R,R'
\to \infty}{\cal S}(0)\right]_{\beta\to \infty} =
\widetilde{E}_c^{(3)}[\ls,R,R'].
\ee

We may use another approach to calculate zeta function at point
$s=-1$ and subtract from integrand all terms of asymptotic
expansion
\bn
\zeta(s,D_2)&=&\widetilde{\zeta}_a(s,D_2) + \widetilde{\zeta}_b(s,D_2)
\label{zetaexptilda}\\
&=& \fr{\sin (\pi s)}{\pi} \sum_{n=-\infty}^{\infty}
\int_{M_n}^\infty dk (k^2 - M_n^2)^{-s} \widetilde{{\cal Z}}_3[k,\ls,R,R']\non\\
&+& \fr{1}{4\pi} \fr 1{\Gamma(s)} \sum_{n=0}^\infty B_{\fr n2}
m^{-2 s + 2 -n}\Gamma(s - 1 + \fr n2).\non
\en
In this case additional contribution appears as \textit{surface}
term in action:
\be\label{energytilde}
{\cal S} = \Omega {\cal S}^{eff}_{\Omega} + \Sigma
\widetilde{{\cal S}}^{eff}_{\Sigma} + \beta \widetilde{E}_c^{(3)}
+ O(e^{-m\beta}),
\ee
where
\bd
\widetilde{{\cal S}}^{eff}_{\Sigma}(0) = {\cal
S}^{eff}_{\Sigma}(0) + E_c^{(3)}[\ls,R\to\infty,R'\to\infty].
\ed
The last term is exactly "surface"\/ contribution $E_{sj}^{(N)}$
given by Eq. \Ref{ENSur}. In the limit $\ls\to \infty$ it is
divergent
\bd
\widetilde{{\cal S}}^{eff}_{\Sigma}(0) \approx \fr{m^3}{12\pi} -
\fr{m^2\ls}{32\pi^2} + \fr{5\ls^3}{576\pi^2} + \fr 12 \Lambda_s  +
\fr{1}{16\pi^2} \left(-\fr{\ls^3}{6} + m^2 \ls\right) \ln\fr\ls m.
\ed

The Lukosz renormalization procedure considered in above sections
gives the same finite result without a renormalization. We will
apply this procedure to action \Ref{S} before renormalization of
constants \Ref{RenConst}. In accordance with this approach we have
to include our system to large box, for example with $x=\pm H$,
where $H> R,\ R'$. The space is divided into three parts: ${\cal
M}_1: \{x\in (-H,-R')\}$, ${\cal M}_2: \{x\in (-R',R)\}$ and
${\cal M}_3: \{x\in (R,H)\}$. The contribution to energy from
boundaries is given by equation
\bd
\triangle {\cal S} = {\cal S}({\cal M}_1)+{\cal S}({\cal M}_2) +
{\cal S}({\cal M}_3) - {\cal S}({\cal M}),
\ed
where the last term is calculated for whole space but without
internal boundaries. In the end we have to tend $H\to\infty$. It
is very easy to apply this formula in our case. One has
\bd
\triangle {\cal S} = 2\beta {\cal S}^{eff}_{\Sigma} (\ls = 0) -
 \fr{\beta}{12\pi^2} \int_m^\infty dx (k^2 -m^2)^{3/2}
\fr\partial{\partial k} \ln \fr{k^2\Psi(k,-R',H)_{\ls=0}
\Psi(k,R,R') \Psi(k,H,-R)_{\ls = 0}}{\Psi(k,H,H)},
\ed
where
\bd
{\cal S}^{eff}_{\Sigma} (\ls = 0) = -\fr{M^3b_{1/2}}{24\pi^{3/2}}
+ \fr 12 \Lambda_s + \fr 12 \ls \varphi_{cl}^2.
\ed
In the limit $H\to\infty$ we obtain the following expression
\bd
\triangle {\cal S} = 2\beta {\cal S}^{eff}_{\Sigma} (\ls = 0) +
\beta \widetilde{E}_c^{(3)}[\ls,R,R']
\ed
and $\widetilde{E}_c^{(3)}[\ls,R,R']$ is given by Eq. \Ref{ENew}.
Therefore we observe that this procedure takes off a volume part,
all pole terms and the renormalization parameter $\mu$. Therefore,
in framework of this renormalization the energy (with dimension
surface energy density) has the following form
\bd
E= \left.\fr{\partial\triangle {\cal S}}{\partial
\beta}\right|_{\beta\to\infty} = -\fr{M^3b_{1/2}}{12\pi^{3/2}}
+ \Lambda_s + \widetilde{E}_c^{(3)}[\ls,R,R'].
\ed
First two terms gives tension of brane with quantum corrections.
The last term completely coincides with that obtained in Sec.
\ref{SecRenormalization} by subtracting all terms of asymptotic
expansion and given by Eq. \Ref{ZOther}.

\section{Conclusion}

Let us summarize our results and observations. In framework of
zeta-function approach we considered three model potentials
namely, single singular potential, step function potential and
three singular potentials. We note problems which were appeared in
connection with these potentials: (i) the Dirichlet limit, $\ls\to
\infty$, is ill defined and the energy is divergent in this limit,
(ii) the energy calculated for step regularized delta-function is
divergent in the sharp limit, $\epsilon\to 0$, when the step
function tends to the delta function, (iii) in massless limit
$m\to 0$ the energy is divergent, too. All of these divergencies
has the same structure -- the energy is logarithmical divergent
with second heat kernel coefficient as a factor (for $3+1$
dimensional case).

In the model of three delta potentials we observe that the energy
of this system for infinite separation of potentials transforms to
a sum of energies each of which is peculiar to single potential
(brane) itself which we called "surface"\/ energy. To obtain
physically reasonable result we suggest to define the Casimir
energy as energy without this "surface"\/ contribution. In this
case the Casimir energy defined in this way  for infinitely
separated potentials (empty space) is zero. The same result may be
obtained by using Lukosz renormalization procedure. We showed
also, that this kind of renormalization correspond to subtracting
from zeta-function all terms of its asymptotic expansion. The
problem appears if we truncate this series and subtract the finite
terms of series. The Casimir energy calculated in this way is well
defined for all physical situations and in the Dirichlet limit the
delta-function transforms to Dirichlet boundary as should be the
case.

Next, we considered in details the effective action for single
delta potential in $(3+1)$ dimensional case in framework of
zeta-regularization approach. To consider renormalization we use
the non-linear $\phi^4$ model with brane part containing the brane
tension and brane's mass. The effective action has surface
contributions except the volume part (effective potential). The
renormalization of the surface part is ambiguous. It is possible
to subtract pole terms by renormalization of the brane parameters.
But there is no universal way to fix these parameters. The
application the Lukosz renormalization procedure takes off all
singularities of the model and gives the Casimir energy which is
coincide with that obtained by suggested procedure.

\section*{Acknowledgments}

The author would like to thank M. Bordag and D. Vassilevich for
reading the manuscript and helpful comments. The author is
grateful to Departamento de F\'{\i}sica, Universidade Federal da
Para\'{\i}ba, Brazil for hospitality. This work was supported in
part by Conselho Nacional de Desenvolvimento Cientifico e
Tecnol\'ogico (CNPq) and by the Russian Foundation for Basic
Research Grants No. 05-02-17344, No. 05-02-39023-SFNS.

\end{document}